\documentclass[10pt,conference]{IEEEtran}
\usepackage{cite}
\usepackage{amsmath,amssymb,amsfonts}
\usepackage{algorithmic}
\usepackage{graphicx}
\usepackage{textcomp}
\usepackage{xcolor}
\usepackage[hyphens]{url}
\usepackage{fancyhdr}
\usepackage{hyperref}
\usepackage{fancyhdr}

\newcommand{\hpcayear}{2025}

\newcommand{\hpcasubmissionnumber}{893}

\title{BOSS: \underline{B}locking algorithm for \underline{o}ptimizing \underline{s}huttling \underline{s}cheduling in Ion Trap}
\def\hpcacameraready{} % Uncomment to build camera-ready version

\newcommand\hpcaauthors{Xian Wu${^\dagger} {^1}$, Chenghong Zhu${^\dagger} {^1}$, Jingbo Wang${^\ddagger}{^*}$, Xin Wang${^\dagger} {^*}$}
\newcommand\hpcaaffiliation{$^\dagger$The Hong Kong University of Science and Technology (Guangzhou) \\$^\ddagger$Beijing Academy of Quantum Information Sciences}
\newcommand\hpcaemail{wangjb@baqis.ac.cn, felixxinwang@hkust-gz.edu.cn}

\usepackage{qcircuit}
\usepackage{amsmath,amsfonts,amssymb,array,graphicx,mathtools,multirow,bm,times,tcolorbox,relsize,booktabs, subfigure}
\usepackage{stfloats}

\newcommand{\nc}{\newcommand}
\nc{\rnc}{\renewcommand}
\nc{\lbar}[1]{\overline{#1}}
\nc{\bra}[1]{\langle#1|}
\nc{\ket}[1]{|#1\rangle}
\nc{\ketbra}[2]{|#1\rangle\!\langle#2|}
\nc{\braket}[2]{\langle#1|#2\rangle}

\nc{\proj}[1]{| #1\rangle\!\langle #1 |}
\nc{\avg}[1]{\langle#1\rangle}
\nc{\rank}{\operatorname{Rank}}
\nc{\smfrac}[2]{\mbox{$\frac{#1}{#2}$}}
\nc{\tr}{\operatorname{Tr}}
\nc{\ox}{\otimes}
\nc{\dg}{\dagger}
\nc{\dn}{\downarrow}
\nc{\cA}{{\cal A}}
\nc{\cB}{{\cal B}}
\nc{\cC}{{\cal C}}
\nc{\cD}{{\cal D}}
\nc{\cE}{{\cal E}}
\nc{\cF}{{\cal F}}
\nc{\cG}{{\cal G}}
\nc{\cH}{{\cal H}}
\nc{\cI}{{\cal I}}
\nc{\cJ}{{\cal J}}
\nc{\cK}{{\cal K}}
\nc{\cL}{{\cal L}}
\nc{\cM}{{\cal M}}
\nc{\cN}{{\cal N}}
\nc{\cO}{{\cal O}}
\nc{\cP}{{\cal P}}
\nc{\cQ}{{\cal Q}}
\nc{\cR}{{\cal R}}
\nc{\cS}{{\cal S}}
\nc{\cT}{{\cal T}}
\nc{\cU}{{\cal U}}
\nc{\cV}{{\cal V}}
\nc{\cX}{{\cal X}}
\nc{\cY}{{\cal Y}}
\nc{\cZ}{{\cal Z}}
\nc{\cW}{{\cal W}}
\nc{\csupp}{{\operatorname{csupp}}}
\nc{\qsupp}{{\operatorname{qsupp}}}
\nc{\var}{{\operatorname{var}}}
\nc{\rar}{\rightarrow}
\nc{\lrar}{\longrightarrow}
\nc{\polylog}{{\operatorname{polylog}}}
\nc{\wt}{{\operatorname{wt}}}
\nc{\av}[1]{{\left\langle {#1} \right\rangle}}
\nc{\supp}{{\operatorname{supp}}}

\nc{\argmin}{{\operatorname{argmin}}}

\nc{\RR}{{{\mathbb R}}}
\nc{\CC}{{{\mathbb C}}}
\nc{\FF}{{{\mathbb F}}}
\nc{\NN}{{{\mathbb N}}}
\nc{\ZZ}{{{\mathbb Z}}}
\nc{\PP}{{{\mathbb P}}}
\nc{\QQ}{{{\mathbb Q}}}
\nc{\UU}{{{\mathbb U}}}
\nc{\EE}{{{\mathbb E}}}
\nc{\id}{{\operatorname{id}}}

\nc{\CHSH}{{\operatorname{CHSH}}}

\usepackage[ruled, vlined, linesnumbered]{algorithm2e}
\author{
  \ifdefined\hpcacameraready
    \IEEEauthorblockN{\hpcaauthors{}}
      \IEEEauthorblockA{
        \hpcaaffiliation{} \\
        \hpcaemail{}
      }
  \else
    \IEEEauthorblockN{\normalsize{HPCA \hpcayear{} Submission
      \textbf{\#\hpcasubmissionnumber{}}} \\
      \IEEEauthorblockA{
        Confidential Draft \\
        Do NOT Distribute!!
      }
    }
  \fi 
}

\fancypagestyle{camerareadyfirstpage}{%
  \fancyhead{}
  
  \fancyhead[C]{
    \ifdefined\aeopen
    \parbox[][12mm][t]{13.5cm}{\hpcayear{} IEEE International Symposium on High-Performance Computer Architecture (HPCA)}    
    \else
      \ifdefined\aereviewed
      \parbox[][12mm][t]{13.5cm}{\hpcayear{} IEEE International Symposium on High-Performance Computer Architecture (HPCA)}
      \else
      \ifdefined\aereproduced
      \parbox[][12mm][t]{13.5cm}{\hpcayear{} IEEE International Symposium on High-Performance Computer Architecture (HPCA)}
      \else
      \parbox[][0mm][t]{13.5cm}{\hpcayear{} IEEE International Symposium on High-Performance Computer Architecture (HPCA)}
    \fi 
    \fi 
    \fi 
    \ifdefined\aeopen 
      \includegraphics[width=12mm,height=12mm]{ae-badges/open-research-objects.pdf}
    \fi 
    \ifdefined\aereviewed
      \includegraphics[width=12mm,height=12mm]{ae-badges/research-objects-reviewed.pdf}
    \fi 
    \ifdefined\aereproduced
      \includegraphics[width=12mm,height=12mm]{ae-badges/results-reproduced.pdf}
    \fi
  }
  \fancyfoot[C]{}
}
\begin{document}
\maketitle

%Enables the camera ready header and footer
\ifdefined\hpcacameraready 
    \ifdefined\notarxiv
      \thispagestyle{camerareadyfirstpage}
      \pagestyle{empty}
    \fi
\else
  \thispagestyle{plain}
  \pagestyle{plain}
\fi

\newcommand{\hpcaheight}{0mm}
\ifdefined\eaopen
\renewcommand{\hpcaheight}{12mm}
\fi

% \end{document}

\begingroup
\renewcommand\thefootnote{1}\footnotetext{Co-first authors.}
\renewcommand\thefootnote{*}\footnotetext{Co-corresponding authors.}
\endgroup
% \begin{document}

%%%%%%%%%%%%%%%%%%%%%%%%%%%%%%%%%%%%%%%%
%%%%%%%% -- PAPER CONTENT STARTS -- %%%%%%%%%

\begin{abstract}
Ion traps stand at the forefront of quantum hardware technology, presenting unparalleled benefits for quantum computing, such as high-fidelity gates, extensive connectivity, and prolonged coherence times. In this context, we explore the critical role of shuttling operations within these systems, especially their influence on the fidelity loss and elongated execution times. To address these challenges, we have developed BOSS, an efficient blocking algorithm tailored to enhance shuttling efficiency. 
This optimization not only bolsters the shuttling process but also elevates the overall efficacy of ion trap devices.
We experimented on multiple applications using two qubit gates up to 4000+ and qubits ranging from 64 to 78. Our method significantly reduces the number of shuttles on most applications, with a maximum reduction of 96.1\%. Additionally, our investigation includes simulations of realistic experimental parameters that incorporate sympathetic cooling, offering a higher fidelity and a refined estimate of execution times that align more closely with practical scenarios.
\end{abstract}

\section{Introduction}

Quantum computing represents one of the most exciting frontiers in modern physics and computer science, promising to revolutionize our computational capabilities \cite{nielsen2010quantum}. Among the various platforms explored for quantum computing, trapped ion technology has emerged as a leading candidate \cite{singer2010colloquium,haffner2008quantum}. It is distinguished by its high degree of qubit control and long coherence times, as well as by its capability for all-to-all qubit connectivity \cite{wang2021single,schafer2018fast,maslov2018outlook}.

Quantum advantage refers to the point at which a quantum computer can perform a calculation or solve a problem that is practically impossible for a classical computer within a reasonable amount of time. For instance, the Shor algorithm provides an exponential speedup over classical algorithms for the prime factoring problem \cite{shor1994algorithms}. The Sycamore circuits were released by Google's quantum computing group in 2019, proving the supremacy of quantum computing  \cite{arute2019quantum}.
In addition, quantum advantages have also been demonstrated in the field of machine learning,
it was proved that quantum machines could learn from exponentially fewer experiments than the number required by conventional experiments \cite{huang2022quantum}. 
Achieving meaningful quantum advantages requires advancements in both quantum hardware and software (algorithms). Consequently, the quantum compilation will be crucial. Similar to classical computing, quantum compilation is the process of converting quantum algorithms into physically executable quantum circuits through a series of operations. A lot of research has been conducted on quantum compilation \cite{khatri2019quantum, ferrari2023modular, gokhale2020optimized}.

Trapped ions offer several significant advantages for quantum computation, yet they consistently encounter scalability challenges \cite{bruzewicz2019trapped}. In particular, trapped ions are subject to the long-range Coulomb interactions that cause the quantized vibrational modes, or phonons, at low temperatures in an ion chain to increase linearly with the addition of each qubit \cite{zhu2006trapped}. These phonons are crucial as they are shared across all qubits in the chain, forming the basis for the full connectivity achievable in ion traps.

To facilitate the interaction between multiple qubits in trapped ion systems, lasers with precisely controlled frequency, amplitude, and phase are employed to individually target corresponding ions \cite{kang2023designing,roos2008ion,green2015phase}. However, this approach faces significant hurdles as the number of ions grows. The frequencies of the phonon modes become densely packed, leading to a notable decline in the fidelity of two-qubit gates as the qubit count increases \cite{monroe2013scaling}. Additionally, the spatial arrangement of ions becomes more complex due to the presence of additional outer qubits. This complexity makes individual ion addressing \cite{mehta2016integrated} a substantial engineering challenge, further complicating the implementation of two-qubit gates that can meet the stringent error thresholds required for fault-tolerant quantum computing \cite{benhelm2008towards}.

In the realm of ion trap technology for quantum computing, laser control is a predominant feature in most hardware platforms. The choice of this control mechanism hinges on whether the hyperfine or fine structure of the ions is utilized to represent qubits \cite{ballance2016high, toyoda2010quantum}. In the current Noisy Intermediate-Scale Quantum (NISQ) era, where lasers are the primary control tools, additional lasers are necessary for cooling and state measurement. Typically, when a Raman configuration is adopted, each qubit necessitates two distinct laser beam paths for quantum gate manipulation: a global laser for overall control and another for independent qubit addressing. This dual-beam setup can be achieved through the use of Acousto-Optic Modulators (AOMs) or Acousto-Optic Deflectors (AODs) \cite{brown2021materials}, which are essential for splitting and directing the optical paths. However, the pursuit of high fidelity in laser-based operations, coupled with the need for sophisticated AOMs and AODs for optical path management, poses substantial challenges. These challenges primarily revolve around the increased volume and cost implications, which significantly impact the scalability of trapped-ion quantum computing systems \cite{mount2016scalable}.

In response to the challenges associated with scaling trapped ion qubits, researchers addressing scalability issues have turned to a key feature of ion trap quantum computing: the movement of qubits in space via microelectrodes \cite{kaushal2020shuttling}. This approach has spurred the development of scalable designs and the ability to implement quantum algorithms on mobile qubits. 
Current advancements in linear trapped ion quantum computing involve segmenting a linear trap into smaller ion chain traps using precise electrode control \cite{jain2020scalable, akhtar2023high}. This method fixes laser beams, eliminating the need for individual adjustments for each quantum operation and allowing ions to be directed to intersect with these beams. This not only streamlines the process but also improves qubit manipulation within the limited quantum decoherence time. Moreover, it stabilizes the number of required lasers, AOMs, and AODs, reducing the costs associated with scaling up ion trap systems.

However, when we optimize for segmented operation of qubits within a linear trap, allowing for their arbitrary exchange and movement, this poses new challenges for efficiently and swiftly compiling general quantum circuits or algorithms to real trapped ion qubits on linear tape trapped-ions quantum hardware platforms. It is crucial to consider how to balance the fidelity improvement from ion trap segmentation, the reduction in hardware costs, and the effective management of the fidelity loss and increased compilation time due to heating effects from ion movement. 
Developing an optimal compilation method for ion trap linear strip traps will provide valuable guidance for implementing algorithms in the NISQ era of trapped-ions quantum computing.

Overall, the main contributions of this work are summarized as follows,
\begin{itemize}
    \item We introduce an innovative blocking algorithm that segments quantum circuits into smaller blocks, allowing for optimized shuttling numbers through efficient strategic scheduling.
    \item  We evaluate our algorithm's performance by estimating execution time and fidelity, incorporating considerations for modern ion trap systems equipped with cooling mechanisms and diverse gate implementation times. 
    \item Compared with existing methods, our algorithm significantly reduces the number of shuttles in most applications, with a maximum reduction of 96.1\%, while the average improvement of shuttle number is 16.6\%.
    It has made significant progress in terms of execution time, with a maximum reduction of 179.6 times and an average improvement of 61.5 times.

\end{itemize}

The structure of the paper is as follows: Section~\ref{sec:background} provides an introduction to the background of quantum computation and an overview of ion trap devices. Section~\ref{sec:simulation} discusses the proposed algorithms for improved shuttling scheduling, along with their complexity. The performance of these methods is evaluated in Section~\ref{sec:evaluation}. Finally, the paper concludes with remarks and insights in Section~\ref{sec:conclusion}.

%%%%%%%%%%%%%%%%%%%%%%%%%%%%%%%%%%%%%%%%%%%%%%%%%%%%%%%%%%%%%%%%%%%%%%%%%%%%%%%%%%%%

\section{Background}\label{sec:background}
In this section, we give a brief overview of quantum computation and relevant background on trapped-ion systems.

\subsection{Quantum Computation}
\paragraph{Qubit} Qubit is the basic unit of quantum information in quantum computing, which has two computational orthogonal basis states $\ket{0}$ and $\ket{1}$.
For a single-qubit state, it can be described by a linear combination of $\ket{0}$ and $\ket{1}$, that is $\ket{\psi} = \alpha\ket{0} + \beta\ket{1}$, where $\alpha$ and $\beta$ are complex amplitudes satisfying $|\alpha|^2 + |\beta|^2 = 1$. As for the $n$ qubit system, it has $2^n$ computational basis states $\ket{0},\ket{1}, \cdots, \ket{2^n-1} $. The $n$ qubit state can be expressed as $\ket{\psi} = \sum_i^{2^n-1} \alpha_i \ket{i}$, where $\alpha_i$ are complex amplitudes satisfying $\sum_i^{2^n-1} |\alpha_i|^2 = 1$.

\paragraph{Quantum operations} Quantum gates are unitary matrices, which transform quantum states via the matrix-vector multiplication. Common single-qubit rotation gates are expressed as $R_a(\theta)=e^{-i\theta \sigma_a/2}$, where $a\in\{x, y, z\}$.
These $\sigma_a$ are Pauli matrices, they can be written as
\begin{equation}
    \sigma_x = \begin{pmatrix}
        0 & 1 \\ 1 & 0
    \end{pmatrix},\quad
    \sigma_y = \begin{pmatrix}
        0 & -i \\ i & 0 \\
    \end{pmatrix},\quad
    \sigma_z = \begin{pmatrix}
        1 & 0 \\ 0 & -1 \\
    \end{pmatrix}.
\end{equation}

\subsection{Dependency Graph}
To describe dependencies in a quantum circuit, we employ a Directed Acyclic Graph (DAG), known as a dependency graph. Each gate in the circuit is represented as a node in this graph, with directed edges indicating dependency relations. For instance, an edge from gate $g_i$ to $g_j$ implies that $g_j$ can only be executed after $g_i$. Typically, we focus on direct predecessor gates. This graph aids in understanding gate relationships and facilitates dividing them into blocks. The longest path in the dependency graph, assuming a uniform time slot, provides a lower bound for the circuit's depth.

For a given gate sequence, one can easily build the dependency graph. It is incredibly straightforward and the time complexity is $O(g)$, where $g$ is the number of input gates. 

\subsection{Trapped-Ion Quantum Computer}
In trapped ions quantum computing, we can manipulate the internal states of ions as qubits. Specifically, to operate on a single qubit, one must select two counter-propagating lasers that are resonant with the frequency of the ion's internal state. By adjusting the duration of the laser pulses and the relative phase difference between the two lasers, we can achieve arbitrary single-qubit operations within the ion trap. The resulting single qubit native gates are \cite{haffner2008quantum}:
\begin{equation}
R(\theta, \phi) 
=
\begin{pmatrix}
\cos(\theta/2) & ie^{i\phi}\sin(\theta/2) \\
ie^{-i\phi}\sin(\theta/2) & \cos(\theta/2)
\end{pmatrix},
\end{equation}
where $\theta$ can be controlled by the pulse time, and $\phi$ can be modulated by the relative phase of the laser.
While in trapped ions quantum computing, the implementation of two-qubit quantum gates requires the use of independently addressed lasers. This can be achieved by modulating the laser beam via AOM or AOD, allowing it to individually address target ions at different positions without causing changes in the internal states of adjacent ions. Additionally, careful modulation of the laser frequency ensures that multiple internal states of ions and phonons are coupled. By designing and modulating pulses through amplitude or phase slicing, we can realize native two-qubit gates on trapped ions quantum computing. The initial concept for two-qubit gates in trapped ions was proposed by Cirac and Zoller \cite{cirac1995quantum}, known as the C-Z gate, which is essentially a controlled phase gate in its matrix form. Subsequently, M{\o}lmer and S{\o}rensen introduced the MS gate, which does not require stringent cooling to the ions ground state, and its matrix form is as follows \cite{sorensen1999quantum}:
\begin{equation}
U_{MS}(\theta_{ij}) = 
\begin{pmatrix}
\cos(\frac{\theta_{ij}}{2}) & 0 & 0 & i\sin(\frac{\theta_{ij}}{2}) \\
0 & \cos(\frac{\theta_{ij}}{2}) & i\sin(\frac{\theta_{ij}}{2}) & 0 \\
0 & i\sin(\frac{\theta_{ij}}{2}) & \cos(\frac{\theta_{ij}}{2}) & 0 \\
i\sin(\frac{\theta_{ij}}{2}) & 0 & 0 & \cos(\frac{\theta_{ij}}{2})
\end{pmatrix},
\end{equation}
where the index $(i,j)$ represents the two ion indexes in a two-qubit gate, and $\theta_{ij}$ represents the entangle angle between two ions.  In the realm of trapped ions quantum computing, the assembly of quantum gates, composed of the native single-qubit gates $R(\theta, \phi)$ and the two-qubit MS gates, establishes a universal set of quantum gates specific to trapped ions. This set enables the decomposition of quantum algorithmic operations into sequences of the aforementioned quantum gates. 
Within the architecture of a linear ion trap, ions are aligned in a sequential, chain-like structure. It is within this configuration that these quantum gates are converted into corresponding laser pulses through a process of compilation. These pulses are then individually addressed at each respective ion. The ions, under the influence of laser beams, undergo efficient internal state transitions. This mechanism is pivotal for the effective processing of quantum information, leveraging the unique dynamics of ion-laser interactions to facilitate quantum computations.

\subsection{TILT Architecture}
\begin{figure}[h]
    \centering
    \includegraphics[width=1.0\linewidth]{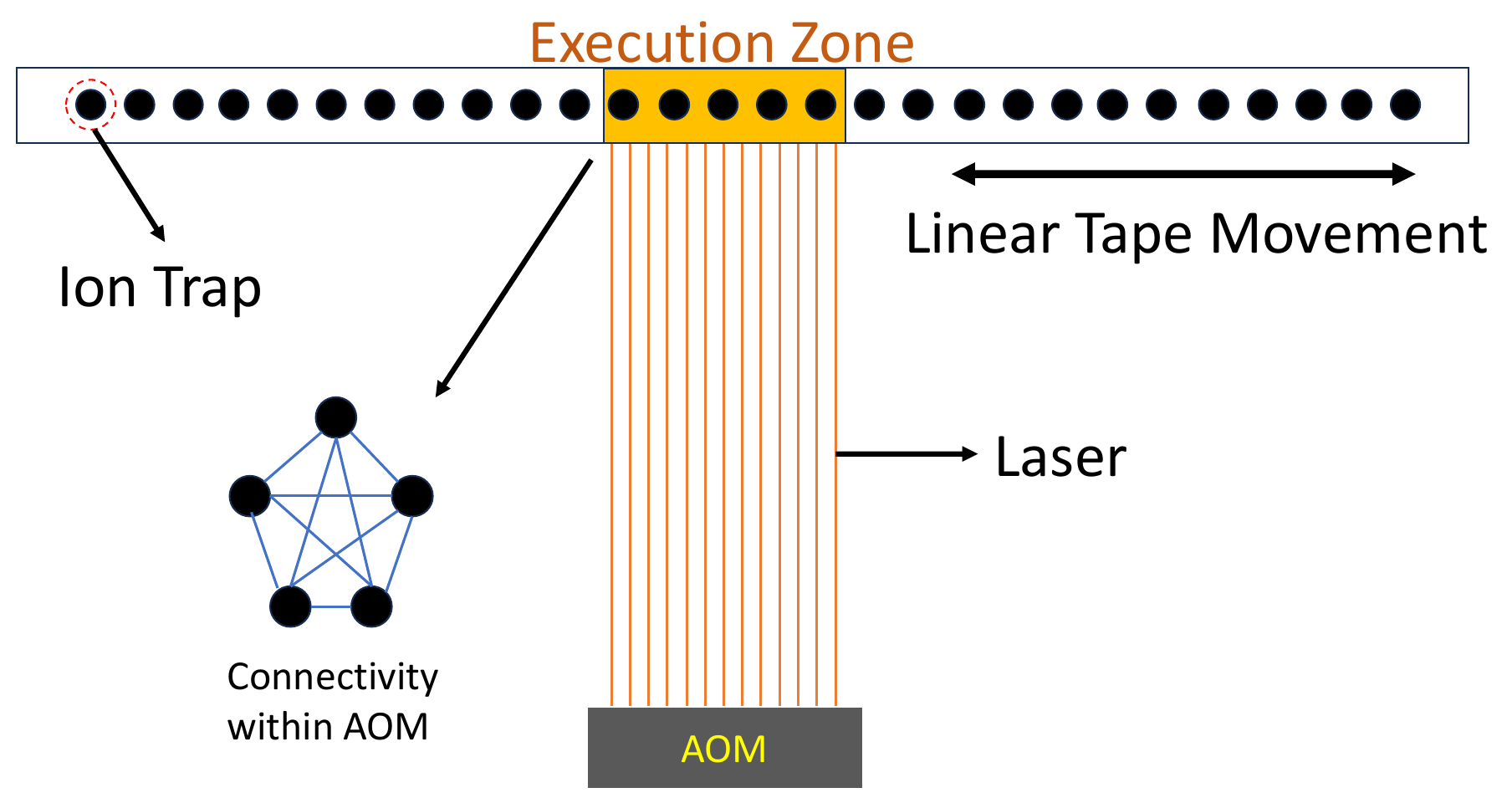}
    \caption{Linear Trapped-Ion quantum computer. Qubits within the Acousto-Optic Modulator (AOM) zone are fully connected. However, to execute quantum operations outside the AOM zone, it becomes necessary to shuttle ions. 
    % \textcolor{blue}{
    In the TILT architecture, each shuttle enables the tape to move any distance.
    % }
    }
    \label{fig:LTI}
\end{figure}
The trapped-ion linear-tape (TILT) was formalized in  \cite{wu2021tilt_iontrap}, which has easier calibration and simplified optics, as well as reduced hardware costs. 
In a TILT system, the entire chain shifts back and forth to operate all qubits, with gates applied only to those qubits that are in the execution zone. Figure~\ref{fig:LTI} provides an overview of TILT.

Another promising trapped-ion architecture is the Quantum Charge Coupled Device (QCCD), which was proposed in \cite{kielpinski2002architecture}.
In contrast to QCCD architectures, which will require more intricate junction shuttle and trap designs as they scale up from their small demonstrations to larger quantum devices, the advantage of the TILT architecture is that it does not require any components that have not previously been demonstrated at the level of complexity necessary for the architecture.
Moreover, gates on TILT devices can be applied simultaneously to one or more qubits \cite{grzesiak2020efficient}. But QCCD is unable to do so at this time. Furthermore, TILT shows how to schedule ions inside a trap.
The QCCD is actually composed of multiple traps, with scheduling within each trap resembling that of TILT. We can schedule large-scale ion trap devices using this strategy in the future.

\subsection{Shuttle}
As depicted in Fig.~\ref{fig:LTI}, the capacity to manipulate qubits simultaneously in ion trap quantum computing is constrained by the availability of laser channels and their corresponding AOM or AOD. 
In an early report from 2018, IonQ revealed their capacity to manage up to 160 qubits within their quantum computing architecture \cite{ionq2018harnesses}. In our research, we follow this established standard, employing a qubit count that remains within this documented limit.
Additionally, single and two-qubit operations are only viable when the ions are positioned within the laser's operational range, an area within the ion trap known as the control region. This raises the important question of how can we address the situation where the number of qubits in the ion trap exceeds the available laser channels.

Under ideal conditions, the internal energy levels of ions or atoms remain unchanged during the shuttle process if it is adiabatic and sufficiently slow, allowing for the movement of atoms without heating. However, practical limitations necessitate the movement of atoms within a finite time, requiring the acceleration of stationary atoms within an electromagnetic field. This acceleration results in the heating of the atoms during finite-time movement.
When the atoms are heated beyond a certain threshold, they escape the trapping potential, leading to qubit loss \cite{reichle2006transport,bluvstein2022quantum}. 
In other words, compared to shuttle distance, the shuttle speed has a greater impact on fidelity.
Essentially, this process can be likened to the accumulation of Pauli Z-type errors during the shuttle.

One conventional approach involves using AOD to adjust the incidence angle of the laser beam \cite{chen2023benchmarking}, thereby modifying the range of ions affected by the beam and enabling control over more ions than there are laser channels. However, the angular range of AOD is restricted, and as the number of qubits in the ion trap increases, achieving global control becomes impractical. An innovative solution to this challenge is to maintain the laser beam's operational area fixed and instead relocate the ions to align with the corresponding positions of the laser beam. This technique, known as "shuttle," capitalizes on the characteristic of ion trap qubits, where the internal state information of the qubits remains unaffected despite the movement of the ions.

The shuttle operation entails designing microelectrodes that, through control feedback, modify the sequence of direct or alternating current applied to them. This generates a spatially dependent dynamic electric potential field above the ion trap chip, prompting the ions to move in response to the changes in the electric potential. This approach significantly enhances the scalability of ion trap quantum computing. The design of microelectrodes is simpler compared to that of laser beams, and the manufacturing cost and technology are more established. Furthermore, the shuttle architecture enables the free movement of ions and the potential division of different functional areas within the ion trap chip, resulting in a more systematic and efficient layout of the quantum chip.

%%%%%%%%%%%%%%%%%%%%%%%%%%%%%%%%%%%%%%%%%%%%%%%%%%%%%%%%%%%%%%%%%%%%%%%%%%%
\section{TILT Simulation}\label{sec:simulation}

\subsection{Overview}

\begin{figure}[h]
    \centering
    \includegraphics[width=\linewidth]{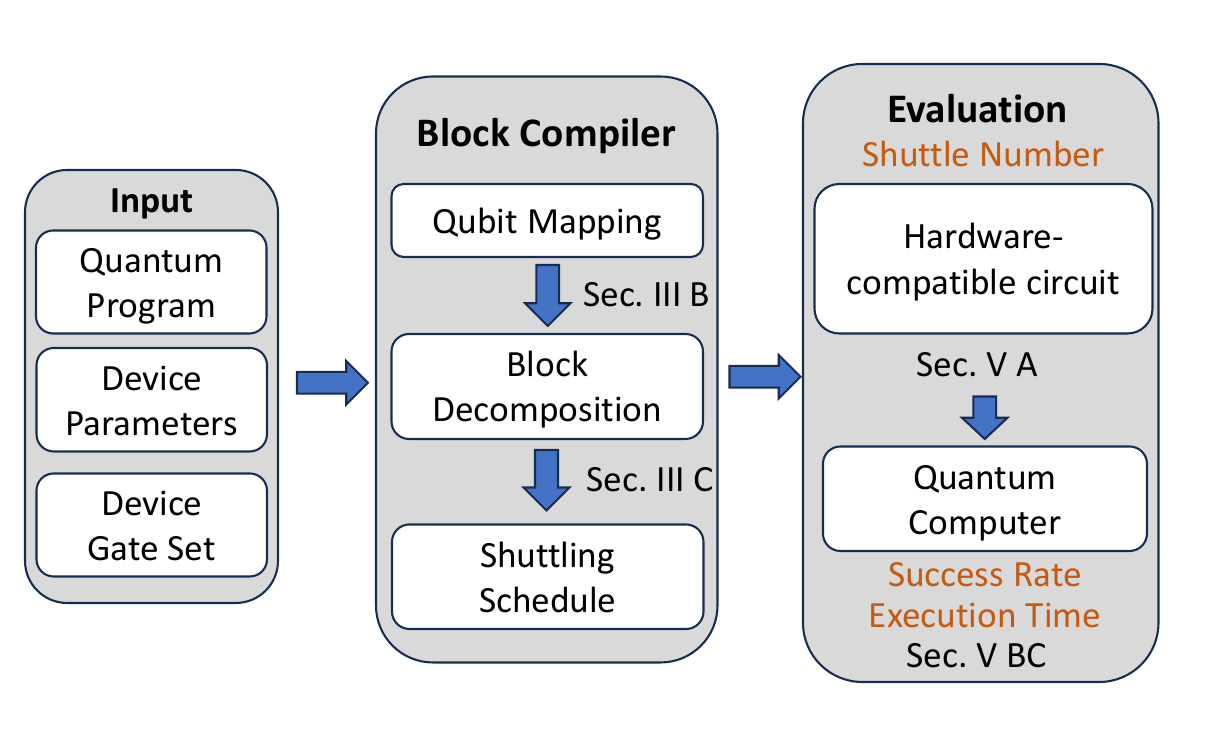}
    \caption{Overview of this work.}
    \label{fig:algorithm_overview}
\end{figure}

In Figure~\ref{fig:algorithm_overview}, it shows the overview of our algorithm. The input of the algorithm consists of a quantum program, device parameters and device native gate sets.  
The algorithm will accept as input the quantum program, parameters of the device and native gate sets of the device.  In the context of linear-type ion trap devices, the device parameters will typically comprise the size of AOM and the number of qubits contained within the trap.  Additionally, the device native gate will comprise the native gates that are customary on ion trap devices.

In section~\ref{sec:algorithm}, we will propose algorithms to minimize the number of shuttling numbers in order to prevent the heating error and achieve a higher success rate for the quantum program.

In section~\ref{sec:evaluation}, the evaluation of the method is conducted through the enumeration of shuttles, evaluation of the success rate and estimation of the execution time.  The noise model that we have taken into account has been used in realistic experiments.

\subsection{Motivation}

The fidelity of a two-qubit gate is intrinsically linked to the total motional energy of the chain during its activation. Shuttling operations impact the success rate of subsequent gates due to the increase in motional energy, which causes gate errors. Despite the implementation of cooling processes, each shuttling operation continues to degrade fidelity. Consequently, it is important to minimize the number of tape moves to maintain high gate fidelity.

Previous work~\cite{wu2021tilt_iontrap} proposes a compiling algorithm to maximize the fidelity by reducing the SWAP gates caused by shuttling operations.
The key idea is to insert swap gates so that each gates can be executed within one execution zone, that is the distance between the qubits affected by each gate is less than a maximum length.
Their algorithm looks at the gate sequence one step at a time. If the distance between the two qubits that a gate affects is larger than the specified value, a heuristic function is used to find the best place to insert extra swap gates. By inserting these swap gates, the distance between the two qubits can be reduced, allowing the original gate to be applied.
After that, they will schedule the tape in a heuristic way. The heuristic function will decide where to move the AOM header if there are no gates in the execution zone that can be executed.

However, their method has several drawbacks. Firstly, to identify suitable locations for inserting swap gates, their heuristic function requires calculating all remaining gates. This approach is time-consuming, as the number of gates escalates rapidly with an increase in qubits. 
Additionally, employing this method may lead to a higher number of shuttling operations. They also did not aim to lower the shuttle count, which could result in a greater fidelity loss. They take care of each gate and ignore the overall information of the circuits.
In this work, we aim to execute a given circuit with a small number of shuttles and a reasonable compiling time.

It is important to highlight the exceptional connectivity of ion trap systems, which are fully connected within a specific area. However, qubits located outside the execution zone are unable to have gates applied to them. To maximize the performance of these devices, it is crucial to leverage this inherent capability. Presently, existing ion trap compiling algorithms predominantly rely on heuristic methods and do not fully exploit the potential of ion trap devices. In response to this challenge, we present a novel algorithm designed to effectively address and optimize this specific issue.

\begin{figure}[h]
    \centering
    \includegraphics[width=\linewidth]{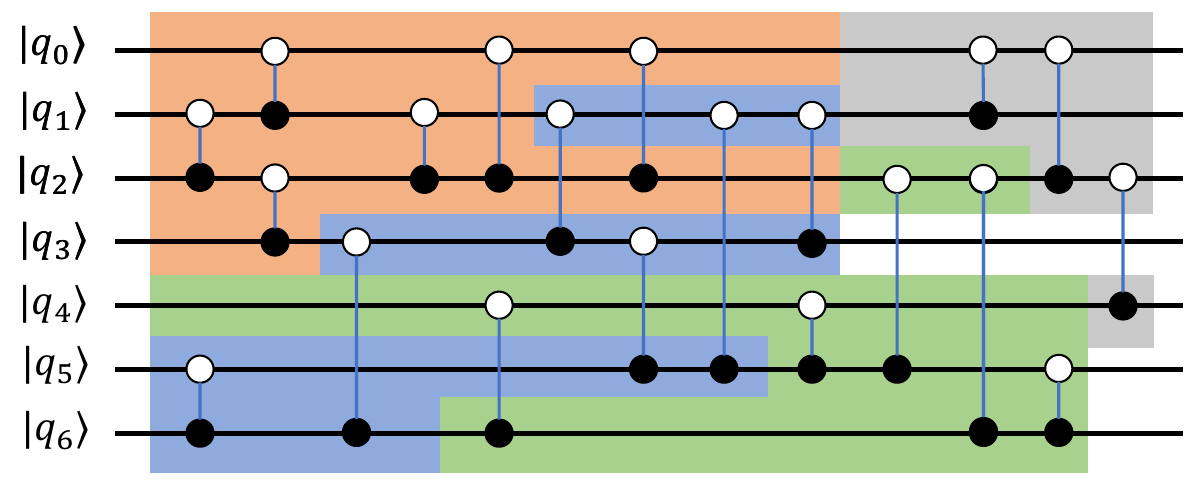}
    \caption{An example of circuit blocking, the maximal block size is 4. Each pair of black and white dots in the figure represents two-qubit gates acting on this qubit pair, and we omit the specific gate representation here.}
    \label{fig:circuit blocking}
\end{figure}

\begin{figure*}[h]
    \centering
    \includegraphics[width=\linewidth]{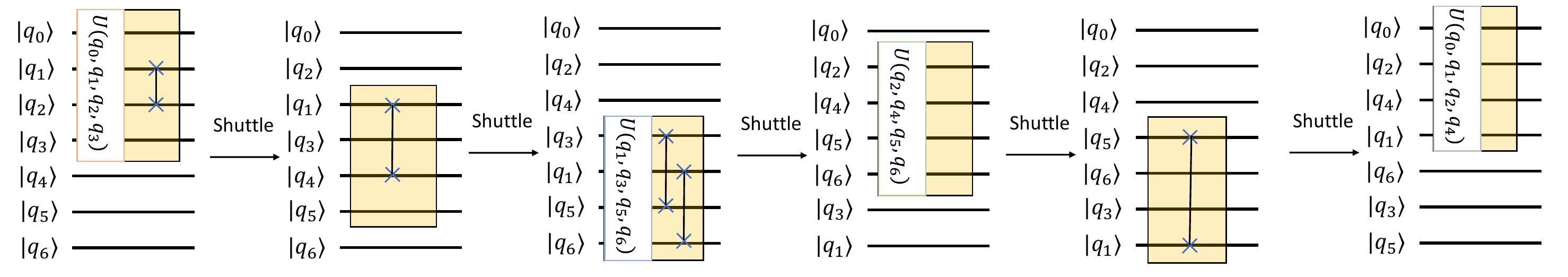}
    \caption{An example of block scheduling. We will execute the block once it has all of the qubits needed inside the execution zone. Otherwise, we will use shuttle and swap gates to bring together the qubits of the requirements.}
    \label{fig:example of block scheduling}
\end{figure*}
The algorithm initiates by partitioning the gates into multiple blocks, with each block operating on no more than $m$ qubits, where $m$ represents the maximum size of the execution zone. The related qubits within each block can then be simultaneously operated within the execution zone. An illustration of this circuit blocking approach is depicted in Figure~\ref{fig:circuit blocking}. For simplicity, single qubit gates have been omitted from the illustration as they can be executed within a block.

Secondly, we need to schedule these blocks. 
The qubits of the following block might not be adjacent when we are executing a block. Therefore, additional shuttle and swap gates should be inserted to ensure the qubits of the next block are adjacent (or they are close enough to share the execution zone together). We call such a procedure block schedule. It is analogous to the superconducting device's gate schedule in certain aspects.
The schedule process following blocking from Fig. \ref{fig:circuit blocking} is shown as an example in Fig. \ref{fig:example of block scheduling}.

These blocks have a dependency relationship and can be thought of as multi-qubit gates. Even though there are fewer blocks than gates, each block requires more qubits than two-qubit gates.

\subsection{Algorithm}\label{sec:algorithm}
\paragraph{Baseline Algorithm}
Here, we begin by introducing a baseline. We first process each executable gate using a simple scheduling technique. 
Until no more gates can be executed, we should select a suitable two-qubit gate and insert a proper swap gate. This algorithm represents a TILT version of a heuristic approach within the context of superconducting quantum systems, which is comparatively straightforward and can leverage the outcomes of related work.
The tape is moved so that it can be executed after the swap gates are inserted.
The pseudocode is shown in Algorithm \ref{baseline}.
\begin{algorithm}[h]
    \label{baseline}
    \caption{TILT mapping (baseline)}

    \textbf{Input} $\mathcal{C}$\\
    $\mathcal{G} = \{\}$\\
    Get initial map $\pi $ and initial execution zone $Z$\\
    \textbf{while} 1 \textbf{do}\\
    \quad $g= get\_next\_gate(\mathcal{C})$\\
    \quad \textbf{if} $g$ is executable \textbf{then}\\
    \quad \quad $\mathcal{G} =\mathcal{G} \cup g$ \\
    \quad \quad update $\mathcal{C}$ \\
    \quad \textbf{else if} $g.q_1$ or $g.q_2$ is in $Z$ \textbf{do}\\    
    \quad \quad $swap$ = \emph{FindProperSwap()}\\
    \quad \quad update mapping $\pi$\\
    \quad \quad $\mathcal{G} =\mathcal{G} \cup swap$ \\
    \quad \textbf{else} \textbf{do}\\
    \quad \quad $tape\_move$ = $FindProperTapeMovement()$\\
    \quad \quad update execution zone $Z$\\
    \quad \quad $\mathcal{G} =\mathcal{G} \cup tape\_move$ \\
    \quad \textbf{end if}\\
    \textbf{end while}\\
    \textbf{return} $\mathcal{G}$
\end{algorithm}

However, the performance of this baseline might not have the desired outcome because it does not fully take advantage of the TILT structure's features.

\paragraph{Detailed Algorithm}

Not every qubit in the execution zone is being used, even though they are all fully connected. Since the distance between any two gates cannot exceed the size of the AOM, the prior approach focuses on inserting swap gates to ensure that all gates can be executed in a single execution zone, which could lead to an excess of shuttles. 
Reducing the number of shuttles is more important than reducing the number of gates, since each shuttle reduces the circuit's overall fidelity.

Executing more gates in each execution zone is our intuition for lowering the shuttle count. We notice that we can gather as many gates as possible if their qubits do not exceed the execution zone size, and we can reduce the vacancy rate of qubits in this way. The vacancy rate indicates the percentage of qubits that are not involved in any gate operations within the execution zone.
In other words, 
our algorithm has the ability to perform more gates between two consecutive shuttle operations.
Furthermore, the shuttle number between two blocks can be bounded, which makes this idea more feasible. 
Therefore, we introduce an algorithm for TILT Blocking.
\begin{algorithm}[t]
    \label{TILT blocking}
    \caption{TILT Blocking}
    \textbf{Input}  circuit $\mathcal{C}$, maximal block size $Z$\\
    Get the dependency graph $\mathcal{G}$ of circuit $\mathcal{C}$\\
    Initialize block list $B=\{\}$\\
    Initialize waiting list $w=\{\}$\\
    Initialize Group list $G$\\
    \textbf{while} $\mathcal{G}.frontier$ is not empty \textbf{do}\\
    \quad Get a gate $g$ from $\mathcal{G}.frontier$\\
    \quad \textbf{if} $g$ is single qubit gate \textbf{do}\\
    \quad \quad $G(g.idx) = G(g.idx) \cup g$\\
    \quad \textbf{else}\\
    \quad \quad $Union\_set=G(g.idx1) \cup G(g.idx2) \cup g$\\
    \quad \quad \textbf{if} $Union\_set$ has no more than $Z$ indexes \textbf{do}\\
    \quad \quad \quad $G(g.idx1) = G(g.idx2) =Union\_set$\\
    \quad \quad\textbf{else}\\
    \quad \quad \quad $w$.append($g$)\\
    \quad \quad \textbf{end if}\\
    \quad \textbf{end if}\\
    \quad \textbf{if} $\mathcal{G}.frontier$ is empty \textbf{do}\\
    \quad \quad Pick a group $p$ from $G$ which has most indexes\\
    \quad \quad \textbf{for} $q\in G$ \textbf{do}\\
    \quad \quad \quad \textbf{if} $q.idx==p.idx$, clear $q$\\
    \quad \quad \textbf{end for}\\
    \quad \quad $B$.append($p$)\\
    \quad \quad $\mathcal{G}.frontier = \mathcal{G}.frontier \cup w$\\
    \quad \quad clear $w$\\
    \quad \textbf{end if}\\
    \textbf{end while}\\
    \textbf{for} $p\in G$ \textbf{do}\\
    \quad $B$.append($p$)\\
    \textbf{end for}\\
    \textbf{return} $B$\\
\end{algorithm}

We first use an algorithm to divide executable circuits into different blocks. Grouping these gates according to their active qubits is the main concept. Two groups are combined if the indexes of a two qubit gate are in both groups and the indexes of the new group do not exceed the maximum group size. The pseudocode is described in Algorithm~\ref{TILT blocking}. 
We can experiment with various strategies in step 7, which will have a significant impact on the partitioning results in this algorithm.

Our work aims at reducing the number of blocks in order to improve operation parallelism. It is anticipated that this reduction will lower the frequency and execution time, enhancing the circuit's overall performance.
Previous works have also reflected this partitioning and scheduling strategy \cite{wu2019ilp_iontrap}, but as the scale increased, the ILP method they used became impractical.
Similar blocking strategies have been found in other studies, including circuit knitting studies \cite{peng2020simulating, gentinetta2024overhead}. 
The main difference with blocking algorithms from circuit cutting, such as CutQC \cite{tang2021cutqc}, is that we do not need to minimize the number of cut edges between subgraphs. This significantly lowers the computational complexity.
In order to facilitate subsequent scheduling, we use a first-in-first-out (FIFO) selection strategy in step 7, which reduces the number of partitioned blocks and increases the frequency of shared qubits among neighboring blocks.
In this case, using FIFO is similar to Breadth-First Search (BFS), with a key benefit being that gates at the same depth are more likely to be grouped into the same block. This helps in the later scheduling and execution of the circuit.

By avoiding the computation of the minimum cut in our setting, we introduce a novel union-find approach for efficient blocking. This method can significantly reduce the vacancy rate of qubits within each block, thereby implying that the same circuit can be accomplished with a reduced number of blocks and potentially decrease the number of shuttles required.
This is the key novelty in our approach.

When we were blocking the circuit, we did not consider the actual qubit mapping on the tape. For a given block, the qubits of it may not be adjacent, we need to introduce extra swap gates and shuttle. 
To get optimal performance, different partitioning strategies must be matched with different scheduling techniques.
Consequently, a wide range of unique strategies can be investigated within this framework. This is left for later work.
We present here a block scheduling method that benefits from limited shuttle movements and low computational complexity.

\begin{figure}[ht]
    \centering
    \includegraphics[width=\linewidth]{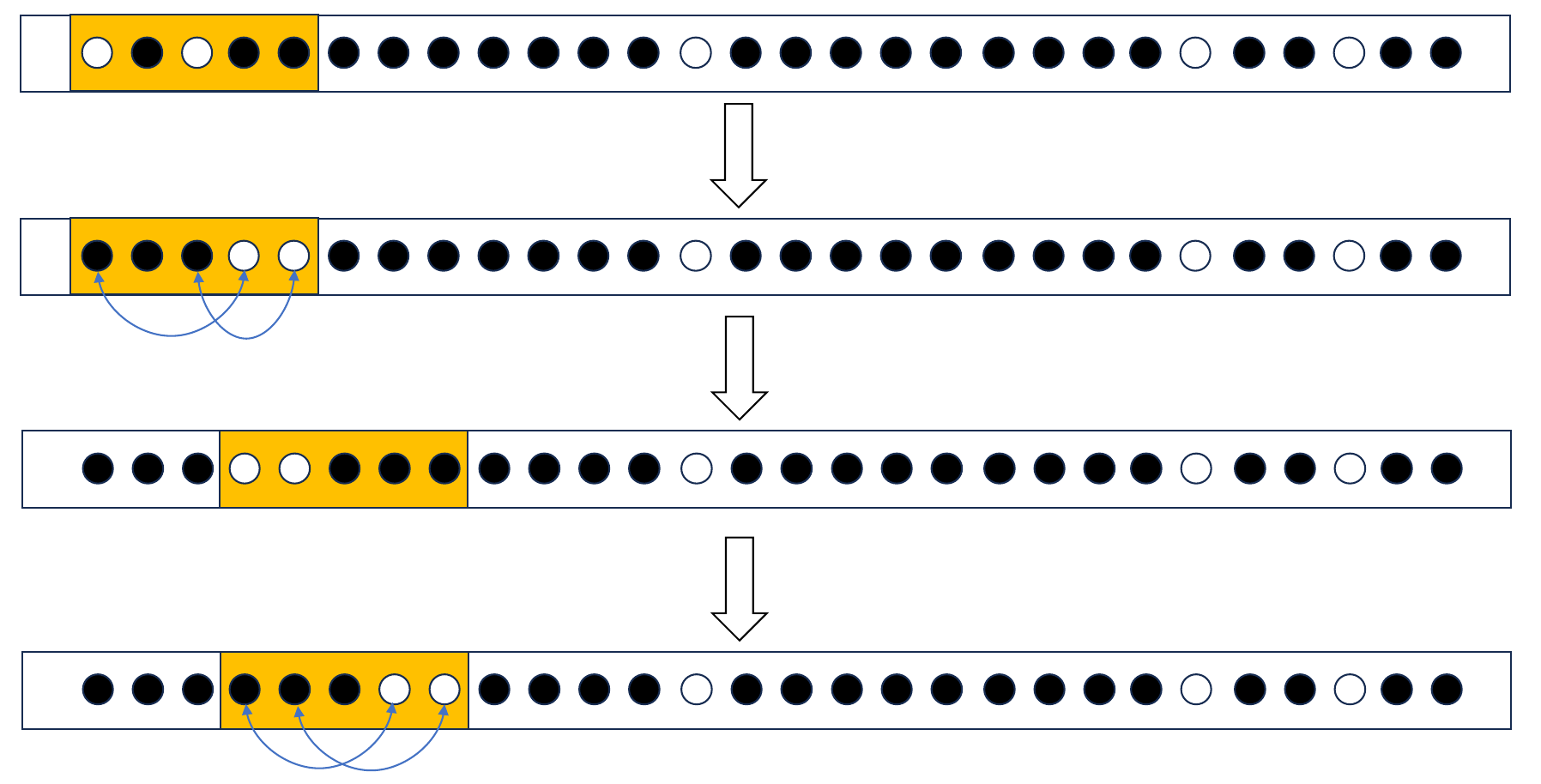}
    \caption{An example of how to move selected qubits. Consider we need to move the white qubits so that they can ultimately be in the same execution zone. We first move the left 2 qubits to the right of the middle one.
    The maximum distance that these two qubits can be moved is $5-2=3$ distance with each shuttle operation.
    Therefore, it would take at least $\lceil d/3 \rceil$ shuttle operations to move them $d$ distance.
    }
    \label{fig:move example}
\end{figure}

For a $m$-size AOM, move $k<m$ qubits $d$ distance only requires $\lceil d/(m-k)\rceil$ shuttles. The minimal shuttle count can be obtained by using $\lfloor k=m/2\rfloor$. 
Such a concept is schematically illustrated in Figure \ref{fig:move example}, where the reader can review additional details.
Therefore, a simple yet effective method is to move the left (right) half qubits to the middle. 
This technique moves the qubits no more than $n$ distance overall because there is less than $n$ distance separating the leftmost and rightmost qubits. 
In addition, it also contains no more than $m/2$ ions for each shuttle since we only move the ion on the left half (or right half).
Thus, the shuttle count between each block is less than $2n/m$.

\begin{algorithm}[h]
    \label{TILT_algo}
    \caption{TILT Block Scheduling}
    \textbf{Input}  block list $B=\{b_0, b_1, \cdots, b_{L-1}\}$, AOM size $Z$\\
    Initialize tape $T=[0,1,\cdots,n-1]$\\
    Initialize execution head $h$\\
    \textbf{for} $b_i\in B$ \textbf{do}\\
    \quad \textbf{if} $\forall idx \in b_i, idx\in [h, h+Z)$ \textbf{do}\\
    \quad \quad execute $b_i$\\
    \quad \quad continue\\
    \quad \textbf{end if}\\
    \quad select the middle index $m_i$ of $b_i$ on the tape $T$\\
    \quad move the left half qubit of $b_i$ to the right of $m_i$\\
    \quad move the right half qubit of $b_i$ to the left of $m_i$\\
    \quad execute $b_i$\\
    \textbf{end for}\\
    
\end{algorithm}
The pseudocode is described in Algorithm~\ref{TILT_algo}.
It is evident that only steps 10 and 11 require additional shuttles, meaning that Algorithm~\ref{TILT_algo} requires no more than $2nL/m$ shuttles.
Figure \ref{fig:move example} provides an example of scheduling qubits to execute a given block within the execution zone.
Because our algorithm can move up to $\lfloor m/2 \rfloor$ qubits in a single shuttle, the swap gates that need to be introduced will not exceed $\lfloor m/2 \rfloor s$, where $s$ is the shuttle number.

\paragraph{Complexity Discussion}
Concerning time complexity, our algorithm shows remarkable effectiveness. The time complexity of the blocking procedure is $O(ng)$, where $g$ denotes the gate number of the input circuit $\mathcal{C}$ and $n$ represents the qubit count. The time complexity of the scheduling procedure does not exceed $O(kmL) = O(nL)$, where $k = \lceil n/m \rceil$. Consequently, the overall time complexity of our algorithm stands at $O(nL + ng)$. Given that $L \leq g/m$, it can equivalently be expressed as $O(ng)$.

The time complexity of previous work \cite{wu2021tilt_iontrap} is relatively high. They first insert swap gates such that the distance of each two-qubit gate is less than the AOM size. They use ha euristic function to find out the best insert swap gates, which requires $O(ng^2)$ in the worst cases, where $n$ is the number of qubits and $g$ is the number of two-qubit gates. The time complexity of its tape movement scheduling is not the dominant term, so the total time complexity is $O(ng^2)$. Our algorithm can always find the solution far more quickly than the previous one.

The worst case scenario for an arbitrary $n$-qubit TILT circuit with a maximal execution zone size of $m$ calls for at least $\Omega(4^{n-m})$ shuttles. Because each $m$-size execution zone introduces a maximum of $4^{m}$ parameters, this is evidently the case.
There are methods based on integer linear programming (ILP) to schedule ion movement and quantum gates in the TILT architecture \cite{wu2019ilp_iontrap}. This ILP approach may be very slow because, as we have already discussed, the shuttle number is exponential to the qubit number in the worst case.

Given that the qubit mapping problem on superconducting devices is an NP-hard problem \cite{siraichi2018qubit}, it makes sense to assume that the gate scheduling problem on ion trap devices is NP-hard as well. 
Therefore, finding a workable solution in a reasonable amount of time is more important than finding the optimal one.

%%%%%%%%%%%%%%%%%%%%%%%%%%%%%%%%%%%%%%%%%%%%%%%%%%%%%%%%%%%%%%%%%%%%%%%%%%%
\section{Experiment Setup}\label{sec:experiment setup}

\subsection{Benchmarks}
Following the best practices outlined in~\cite{wu2021tilt_iontrap}, we assess the efficacy of blocking shuttling through the utilization of several significant quantum applications as benchmark cases.

Table.~\ref{table:benchmarks} in our study provides a comprehensive overview of various application circuits considered in our research. These applications range from circuits designed with nearest-neighbor gates and short-distance patterns, which are typically suited for near-term superconducting devices, to well-known algorithms that demonstrate quantum advantages. This diverse selection of applications serves to showcase the versatility and wide-ranging applicability of our proposed methods in different quantum computing contexts.

\begin{table}[h]
\centering
 \begin{tabular}{|c  c  c  c|} 
 \hline
 \textbf{Application} & \textbf{Qubits} & \textbf{2Q Gates }& \textbf{Communication} \\
 \hline
 \hline
 Adder & 66 & 545 & Short-distance gates  \\ 
 BV & 65 & 64 & Long-distance gates \\
 QAOA & 64 & 1260 & Nearest-neighbor gates \\
ALT&64&1260&Nearest-neighbor gates \\
 RCS & 64 & 560 & Nearest-neighbor gates \\
 QFT & 64 & 4032 & Long-distance gates \\
 SQRT & 78 & 1028 & Long-distance gates \\
 % \hline
 \hline
\end{tabular}
\caption{List of benchmarks.}
\label{table:benchmarks}
\end{table}

In detail, the adder benchmark is based on the Cucarro adder~\cite{cuccaro2004new}, which is a 32-bit adder and we get its qasm from~\cite{quantum-circuit-generator}. Bernstein-Vazirani (BV) is a NISQ application commonly used to benchmark devices.  Quantum Approximate Optimization Algorithm (QAOA) is a hybrid iterative method for solving combinatorial optimization problems, i.e., maxcut~\cite{farhi2014quantum}, SK model~\cite{farhi2022quantum}.
Furthermore, we consider alternating layered ansatz (ALT)~\cite{farhi2014quantum}, which is commonly used in variational quantum eigensolver (VQE).
Random Circuit Sampling (RCS) was proposed by Google to show quantum supremacy, and we obtain its qasm from~\cite{google2023cirq}. Quantum Fourier Transform (QFT)~\cite{bernstein1993quantum} is an key ingredient for Shor's algorithm~\cite{shor1994algorithms} and phase estimation algorithms~\cite{kitaev2002classical}.
The square root (SQRT) application is from ScaffCC~\cite{javadiabhari2015scaffcc}, and it finds the square root number by using Grover's search.

In our analysis, we focus exclusively on two-qubit gates due to their longer runtimes and higher error rates compared to single-qubit gates. We note that single-qubit gates can be merged to limit their number to at most two between any two-qubit gates, effectively reducing their presence. However, for consistency and to facilitate comparisons with prior work~\cite{wu2021tilt_iontrap}, we exclude single-qubit gates from our considerations.

\subsection{Simulation Parameters}
We assess the efficacy of the methodologies using various circuits and AOM sizes of 16 and 32. All experiments are performed on a Windows 11 system with AMD Ryzen 5 6600H CPU and 16-GB physical memory using Python 3.9.

%%%%%%%%%%%%%%%%%%%%%%%%%%%%%%%%%%%%%%%%%%%%%%%%%%%%%%%%%%%%%%%%%%%%%%%%%%%
\section{Evaluation}\label{sec:evaluation}
% \textcolor{blue}{1}

\subsection{Shuttling Number}

\begin{figure}[h]
    \centering
    \includegraphics[width=\linewidth]{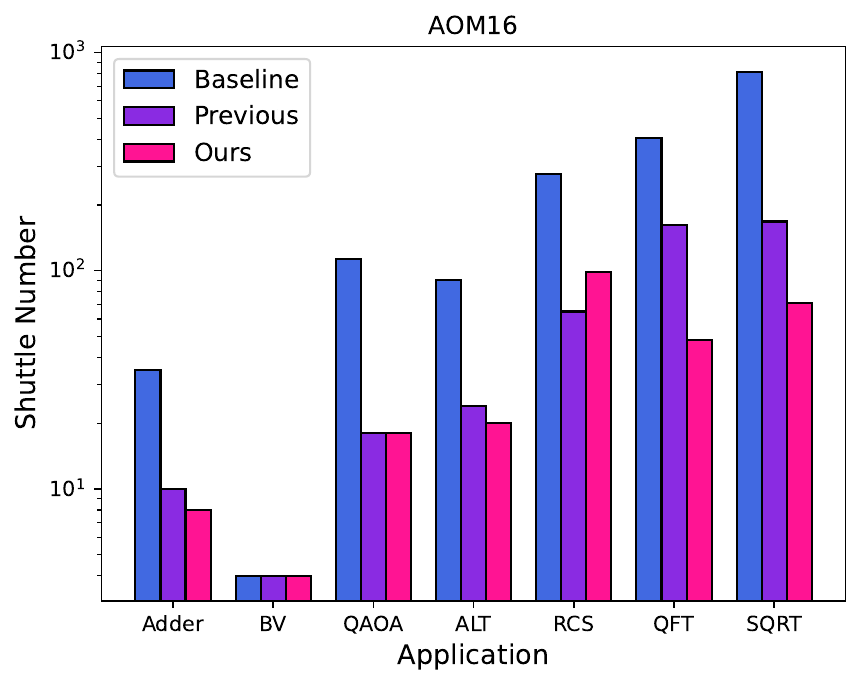}
    \caption{Number of shuttling operations (Lower the better).} 
    \label{fig:shuttle number}
\end{figure}

To evaluate the effectiveness of our method, we use previously mentioned applications for benchmarking. Figure \ref{fig:shuttle number} shows the shuttle number of different applications when AOM size is 16, and our method has shown improvement on most applications.

\begin{table*}[b]
\centering
\begin{tabular}{|c|c|c|c|c|c|c|c|c|c|c|c|}
\hline
 \textbf{Application} & \textbf{AOM size}
 & $T_{pre}(s)$ & $T_{boss}(s)$&$S_{base}$& $S_{pre}$ & $S_{boss}$ & $\Delta S$ & $\Delta S/S_{pre}$ 
 &$t_{pre}(s)$ &$t_{boss}(s)$ &$t_{pre}/t_{boss}$ \\
\hline
\hline
 Adder&16 &6.216& 0.007&35& 10 & 8  & 2& 20\% 
 & 2.967 & 0.0378 & 78.6\\
BV&16  &1.579&0.006 & 4 & 4& 4&0 & 0\% 
&0.856& 0.0354 & 24.2\\
QAOA&16  &2.761 &0.038 &113 & 18 & 18 &0 & 0\% 
& 1.564 & 0.0418 & 37.4 \\
ALT &16 &2.880&0.013 &91 &24& 20& 4 &16.7\% &1.311 &0.0256 &51.2\\
 RCS&16  & 3.149&0.045&277&65  & 106 &-41 & -63.1\% 
 &1.704 & 0.2057 & 8.3 \\
 QFT &16 &64.823& 0.064 &407 & 162 & 48  &114 & 70.4\%  
 & 24.820 & 0.4405 & 56.3\\
 SQRT&16  &131.245 & 0.430 & 816 & 168 & 71 &97  & 57.7\%  
 & 46.554 & 0.2593 & 179.6 \\
 \hline
 Adder&32&3.250&0.006 & 5 & 5& 4 &1 & 20\% 
 &3.252 & 0.0372 & 87.5 \\
BV&32 &0.902& 0.005& 2& 2 & 2&0 & 0\% 
&0.987 & 0.0527 & 18.7\\
QAOA&32   &1.112&0.029 & 40 & 4&  4&0  & 0\%
& 1.357 & 0.0284 & 47.7\\
ALT &32&1.404&0.019 &38 &8 &5 &3& 37.5\% & 1.017  & 0.0178  & 57.1 \\
 RCS&32 &0.681&0.017 & 86 & 11 &  21&-10  & -90.9\% 
 &0.856 &0.1307 & 6.5\\
 QFT &32 &37.341& 0.051& 69 & 69& 8 & 61 & 88.4\% 
 & 33.876 & 0.3926 & 86.3\\
 SQRT&32  &56.309 & 0.013 & 431 & 76 & 3 & 73 & 96.1\%  
 & 40.817 & 0.2070 & 107.1\\
\hline
\end{tabular}
% }
\caption{Comparison with previous work~\cite{wu2021tilt_iontrap} on number of compilation time, shuttle operations and time evaluation (Lower is better).}
$T_{pre}$:
compilation time of previous.
$T_{boss}$:
compilation time of BOSS method.
$S_{pre}$:
shuttle number of previous.
$S_{base}$: shuttle number of baseline. 

$s_{boss}$: shuttle number of BOSS method. $\Delta S:=S_{pre} - S_{boss}$.
$t_{pre}$: execution time of previous (Trout model).

$t_{boss}$: execution time of BOSS method (Trout model).

\label{table:shuttle_num_compare}
\end{table*}

More details can be found in the Table.~\ref{table:shuttle_num_compare}, it presents a comprehensive comparison with previous work~\cite{wu2021tilt_iontrap}. 
The experimental results show that in most applications, our algorithm exhibits fewer shuttles. 
Our approach reduces QFT and SQRT significantly when AOM size is 32, by 88.4\% and 96.1\%, respectively. 
Furthermore, our algorithm operates at a much faster pace than previous approaches, suggesting that our approach has greater potential for more intricate circuits. Compared to previous methods, our approach uses more swap gates—for instance, our method on AOM16's QFT requires 336 swap gates, whereas the previous method only needed about 120 swap gates.
But the difference in number is negligible when you consider all of the gates.

Our algorithm performs worse than existing algorithms on RCS. This is mainly due to the inherent characteristics of circuits like RCS, where the distance between the qubits of all two bit gates does not exceed 15 under trivial initial mappings.
Therefore, swap insertion is not required to be taken into account by the previous algorithm.
In this unique circumstance, they have a significant advantage because they only need to think about the shuttle schedule.

Compared to existing algorithms, our method better utilizes the execution zone.
The algorithm in previous work~\cite{wu2021tilt_iontrap} only reduces the distance between two-qubit gates, but does not make good use of the feature that only a specific area on the tape can execute gates at the same time.
Our algorithm divides the circuit into blocks so that more gates can be executed within the same block. 
Our algorithm can benefit from reducing the vacancy rate.
Take QFT as an example, our algorithm ensures that no qubit within the execution zone remains idle, which allows us to perform more gates between two consecutive shuttle operations in comparison to the previous techniques~\cite{wu2021tilt_iontrap}.

\subsection{Execution time}

Estimating the execution time of quantum circuits on ion trap systems is pivotal in providing valuable guidance to physicists involved in the development of long-coherence quantum computers. By understanding the time required for circuit execution, researchers can gain insights into the practicality of various applications on quantum computers. If the estimated execution time is excessively prolonged, it may indicate limitations that could impact the practicality of certain quantum computing applications. In this context, our work explores diverse implementation options for the two-qubit gate.

\begin{figure*}[b]
    \centering
    \subfigure{
        \includegraphics[width=0.45\linewidth]{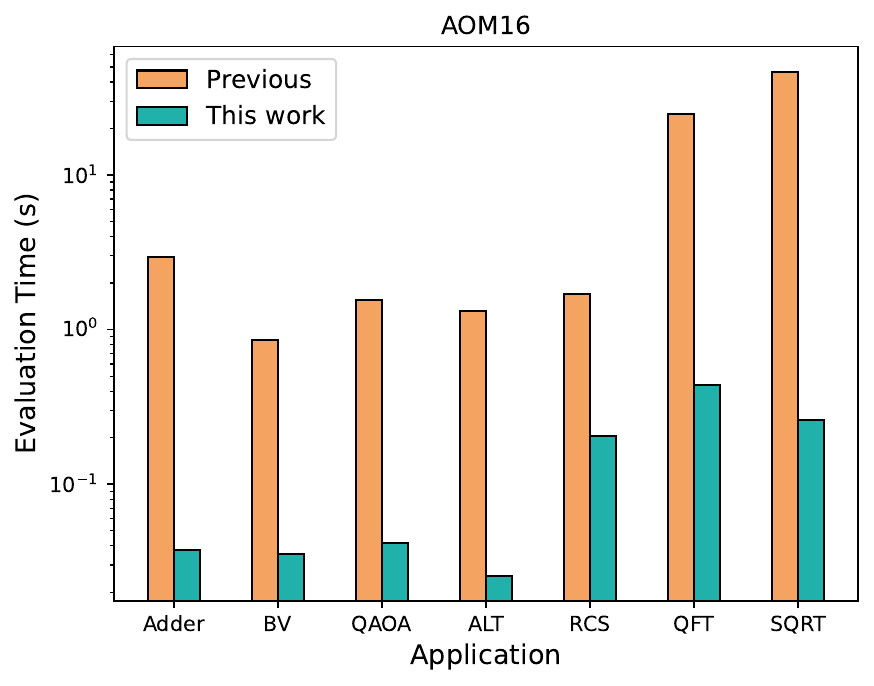}
    }
    \subfigure{
        \includegraphics[width=0.45\linewidth]{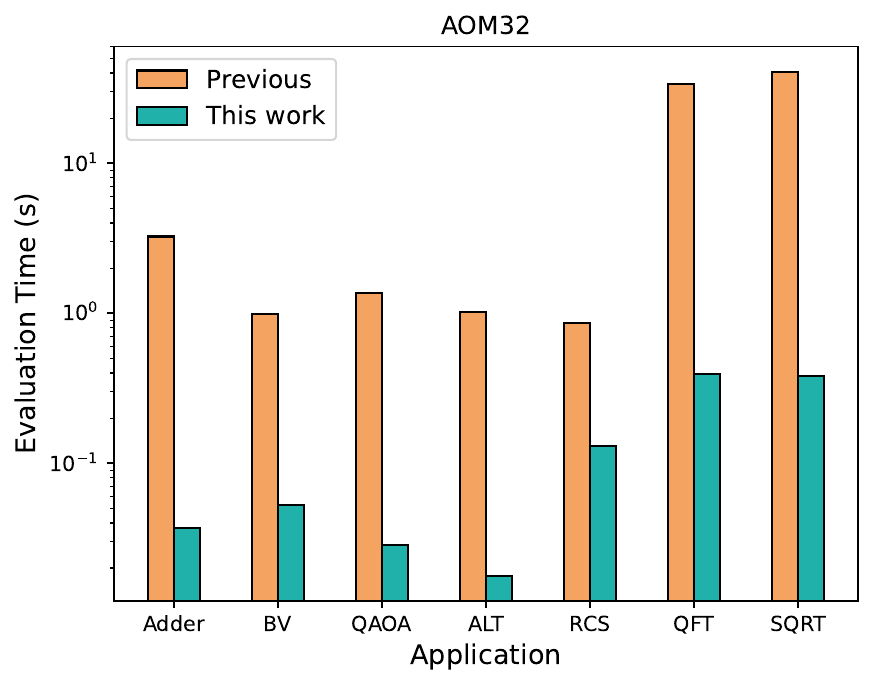}
    }
    \caption{ A comparative analysis of the execution time with previous work~\cite{wu2021tilt_iontrap} utilizing AM gate implementation~\cite{Trout2018AM}. }
    \label{fig:time comparison}
\end{figure*}
\paragraph{Two-qubit Gate Implementation Time}In the simulation, we consider amplitude modulated gates~\cite{wu2018noise, Choi2014AM, Trout2018AM} and phase modulated gates~\cite{milne2020phase} for two-qubit gate implementation.  Both gate implementation time of AM and PM gates are proportional to the distance of ions. For the first type of AM gate implementation~\cite{wu2018noise}, the time for implementation is,
\begin{equation}
    \tau_{\text{Duan}}(d) = 100d - 22.
\end{equation}
For another faster but less accurate AM gates~\cite{Trout2018AM},
the time for implementation is,
\begin{equation}
    \tau_{\text{Trout}}(d) = 38d + 10.
\end{equation}
The PM gates implementation has weaker dependence on distance but slower for nearby qubits~\cite{milne2020phase}, which has implementation time as,
\begin{equation}
    \tau_{\text{PM}}(d) = 5d + 160,
\end{equation}
where $d$ is the distance between two ions and the time is given in microseconds.

\paragraph{Sympathetic Cooling Time}
In the domain of trapped-ion quantum computing, lasers serve a multifaceted purpose, being instrumental not only in executing standard quantum gate operations but also in facilitating critical processes such as ion cooling, initial state preparation, and quantum state readout. The parameters of these laser pulses, including frequency and duration, are meticulously customized to suit the specific requirements of each individual operation. It is the temporal aspect of these additional operations that warrant our attention, particularly with respect to estimating the execution time involved.

The initial states cooling of ions commonly commences with Doppler cooling, a technique wherein a laser is utilized to diminish the vibrational energy of the ions \cite{wineland1979laser}. This initial cooling stage conventionally takes approximately 10 ms. Subsequent to the Doppler cooling, sideband cooling is frequently implemented as an additional measure to attenuate thermal fluctuations, thereby bringing the ions nearer to their quantum ground state. This secondary cooling process typically requires about 50 µs \cite{chen2020efficient}. To represent the aggregate duration required for the ion preparation, we designate the total initial cooling time as 
\begin{equation}
    t_1 = 10 \text{ ms} + 50 \text{ µs}.
\end{equation}

\begin{figure*}[h]
    \centering
    \subfigure{
        \includegraphics[width=0.45\linewidth]{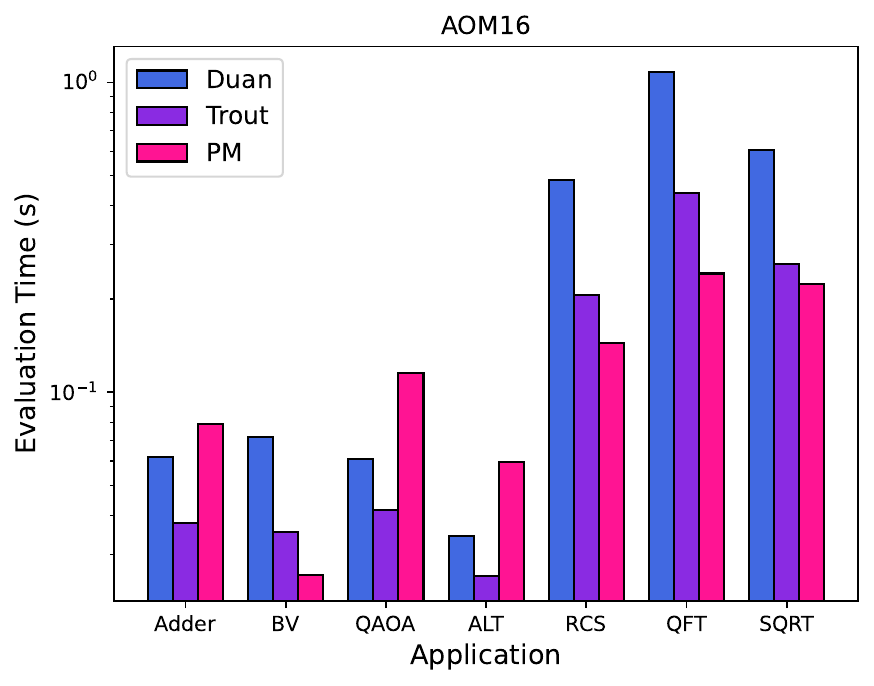}
    }
    \subfigure{
        \includegraphics[width=0.45\linewidth]{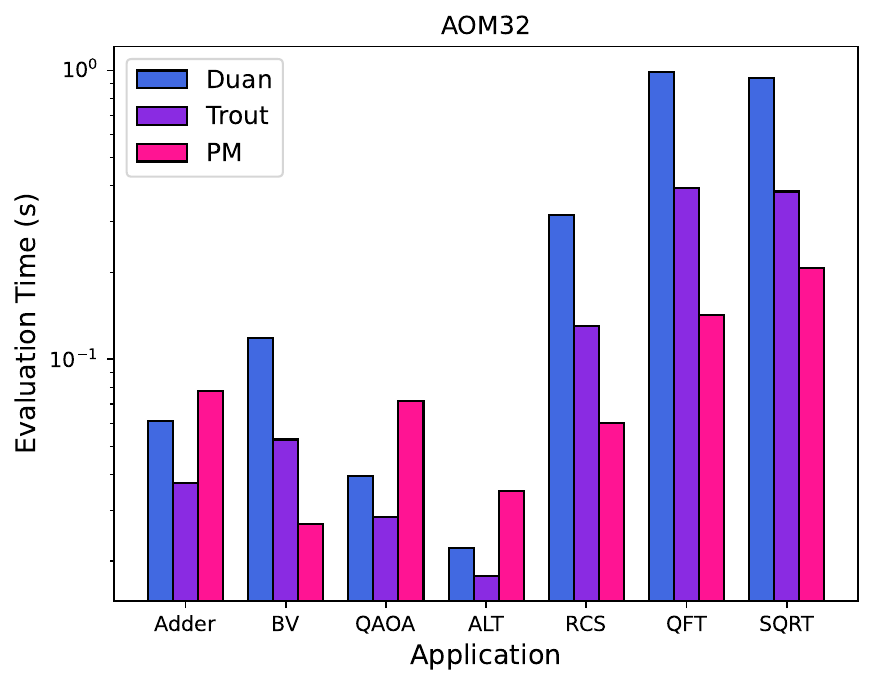}
    }
    \caption{Time evaluation on different applications with Duan, Trout AM gate and PM gate implementation. Circuits with a high percentage of short-distance gates work better for AM, while applications with long-distance gates work better for PM.}
    \label{fig:time evaluation}
\end{figure*}
Upon completion of the initial state preparation, it is observed that the execution of a quantum circuit can lead to the gradual heating of the ion chain, particularly evident during shuttle operations involving ion movement. To mitigate the adverse impact of heat-induced quantum noise, cooling operations can be integrated following each shuttle operation. The duration required for this cooling process is akin to that of sideband cooling, with a typical reference timeframe of approximately 40 µs \cite{labaziewicz2008suppression}.
In this paper, we use $t_2$ to represent the midway cooling time,
\begin{equation}
    t_2 = 40 \text{ µs} \times s_m,
\end{equation}
where $s_m$ is the number of shuttle operations.

At the culmination of all quantum circuit operations, a vital step involves the readout process, which is essential for extracting quantum information from the ion trap system. In ion trap systems, it is important to note that the fidelity of the readout is directly correlated with the duration of the readout process. In order to achieve a readout fidelity over 0.9999, the readout time typically exceeds 150 $\text{ µs}$~\cite{myerson2008high}. This delicate balance between readout duration and fidelity is a crucial aspect that directly influences the efficiency and accuracy of ion trap quantum computing systems.
We use $t_3=150 \text{ µs}$ to represent the readout time.

Combining all parameters, we evaluate the execution time by analyzing the compiled circuit depth, the required shuttle move distance and the cooling time as follows,
\begin{equation}
    t_{exec} = v_m \times dist + \sum_d t_d +t_1+t_2 + t_3,
\end{equation}
where $t_d$ is the maximum gate time for the circuit depth $d$, $v_m = 1\text{ µm}/\text{ µs}$ is the shuttling speed~\cite{Bruzewicz2019} and $dist$ is the total shuttle distance.

Figure~\ref{fig:time comparison} presents a comparative analysis of our work with previous research in terms of total execution time~\cite{wu2021tilt_iontrap}. 
The results in the figure indicate that our algorithm has shown significant progress in execution time, with the highest performance being 179 times on the SQRT of AOM 16.
The improved execution time is primarily attributed to the reduced idle rate among qubits in the execution zone, which boosts the concurrent execution efficiency of two-qubit gates, in addition to the optimized preprocessing time.

By implementing the Trout AM gate, we have been able to further reduce execution time, projecting that applications could potentially conclude within a matter of seconds. 
Furthermore, our evaluation encompasses an in-depth study of ion cooling, initial state preparation, and readout timing, thereby offering a more meticulous estimation of execution timing for contemporary ion trap systems. Given that trapped-ion qubits exhibit extended coherence times in comparison to other quantum hardware~\cite{wang2021single}, our findings underscore the viability of ion trap systems for executing comprehensive applications with profound depth circuits.

In Figure~\ref{fig:time evaluation}, we extend our analysis to include a comparison between the execution times associated with both AM and PM gate implementations across various applications. Our results demonstrate that, in the context of short-distance communication circuits, implementations utilizing the Trout gate typically yield reduced execution times. Conversely, for long-distance communication circuits, exemplified by QFT and SQRT, the PM gate implementation may offer superior performance in terms of execution time due to its diminished reliance on qubit-qubit separation.

Because AOM 32 permits gates to act on farther qubits, we will discover that in some applications, it takes longer time for AOM 32 to execute than AOM16.  We don't focus on how to reduce the execution time here, optimizing the execution time could be a future work.

\subsection{Fidelity}

\paragraph{Fidelity Estimation. }
In trapped-ion quantum computing, the fidelity of single-qubit gates is often remarkably high. These gates solely involve transitions within the internal states of the ions and do not entangle with the phonon modes. Consequently, in our subsequent fidelity estimations for quantum algorithms, the impact of single-qubit gates is not considered. However, for two-qubit gates, the situation is more complex. Due to the full connectivity of qubits in the laser operation area, the fidelity of two-qubit gates is intrinsically linked to the number of ions in the operational region. This is because it involves entanglement exchanges between ions and phonons. Assuming the use of amplitude-modulated MS gates, the pulse duration and the number of pulse slices required for more phonons are positively correlated. Generally, the key parameters of the two-qubit gates, entangle angle $\theta_{ij}$ have a quadratic relationship, approximated with pulse amplitude as $\theta_{ij}\propto \Omega^2$. Consequently, we can estimate the fidelity of two-qubit gates as 
\begin{equation}
    \label{gate_fid}
    F_{gate} = 1 - \epsilon_{laser} N^2, 
\end{equation}
where N is the number of ions in the laser interaction region, and $\epsilon_{laser}$ represents the precision coefficient of instrument manipulation~\cite{ gale2020optimized, martinez2022analytical,  wu2018noise}. 
In our simulation, we set $\epsilon_{laser}=1/256000$, ensuring that two qubit gates achieve a fidelity of 99.9\% in AOM 16 cases.

Moving ions does not, in theory, change the quantum information that is carried by their internal states when it comes to shuttle operations.
This is especially true after the shuttle operation has effectively suppressed any heating effects by re-cooling the ion chain.
Therefore, we do not need to consider fidelity loss due to heating effects. However, the entanglement between ions and phonons is not resolved by moving the ions, and errors accumulate with each step of the shuttle operation. 
We can estimate the loss of fidelity in the quantum circuit due to the $m$-th shuttle operations using the following formula,
\begin{equation}
F_{\text{shuttle}}=1-\epsilon_{\text{shuttle}}m,    
\end{equation}
where $\epsilon_{shuttle}$ is the residual entangle between ion and phonon during the shuttle operation.
In our simulation, we set $\epsilon_{\text{shuttle}}=0.001$. 
It is important to note that the fidelity of a single shuttle operation decreases linearly with the number of times it is performed. Quantitatively, the fidelity loss incurred by the $m$-th shuttle is greater than the loss caused by $m$ two-qubit gates.

\begin{figure}[h]
    \centering
    \includegraphics[width=\linewidth]{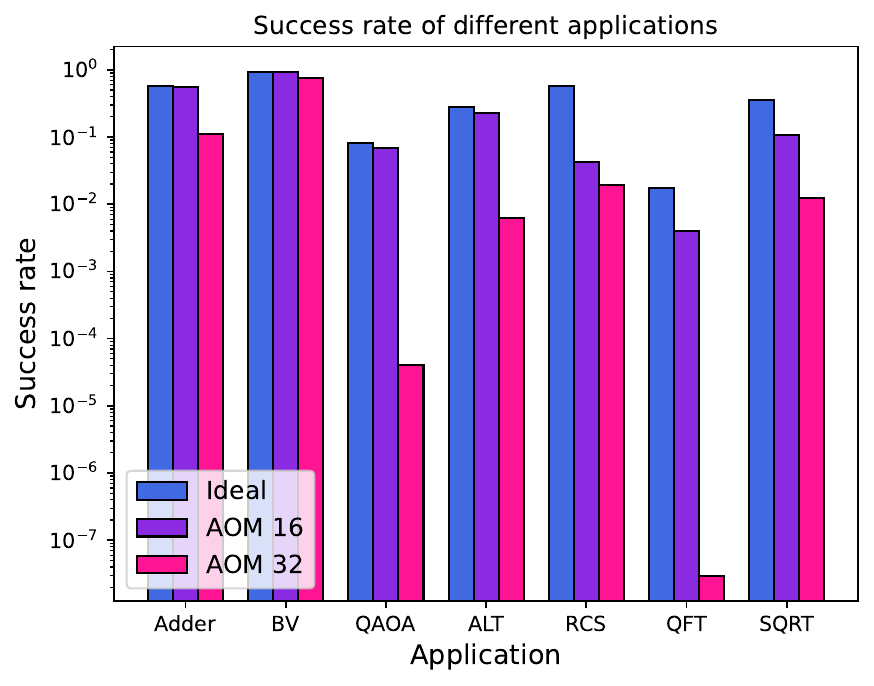}
    \caption{Success rate comparison on different applications (Higher the better).} 
    \label{fig:success rate comparison}
\end{figure}

\paragraph{Success Rate Evaluation. }
As a result, the following formula is used to evaluate the overall success rate. 
\begin{equation}
    F =  F_{gate} ^{g}\times \prod_{m=1}^S(1-\epsilon_{\text{shuttle}}m) 
\end{equation}
Where $S$ represents the total shuttle number and the total gate number is indicated by the $g$.
In this part, we will discuss the relationship between success rate and AOM size. 
The experiment result is shown as Fig \ref{fig:success rate comparison}.

We have a higher success rate than in previous work \cite{wu2021tilt_iontrap} because we make good use of the cooling process. 
For example, their success rate is less than $1\times10^{-14}$ for QFT with an AOM size of 16, whereas ours exceeds $4\times10^{-3}$.
An ideal device means it has sufficient laser control for each quantum bit, which can also be understood as AOM size equal to the number of qubits. 
That is to say, two-qubit gates can be applied to any pair of qubits so that extra swap gates and shuttles become unnecessary. Moreover, the gate fidelity of the ideal model will not decrease as the AOM size grows.
This ideal model is consistent with the one in previous work \cite{wu2021tilt_iontrap}.

Our gate fidelity model assumes gate fidelity decreases rapidly as the AOM size increases, which results in the overall fidelity reduction as the AOM size increases.
This idea is supported by the experimental results shown in Figure \ref{fig:success rate comparison}, which are consistent with what we had expected.
There is a slight divergence between the results and the previous research \cite{wu2021tilt_iontrap}, which can be attributed to the different formulas we used to estimate gate fidelity. The equation we used is more accurate in reflecting the real conditions, so it is more useful as a reference.
Moreover, as technology advances, it will be more and more clear how much better our method is. Our approach, which uses fewer shuttles, will perform better when the fidelity of the gate does not drop off quickly as the AOM size increases.

\subsection{Scalability}
\begin{figure}[h]
    \centering
    \includegraphics[width=1.0\linewidth]{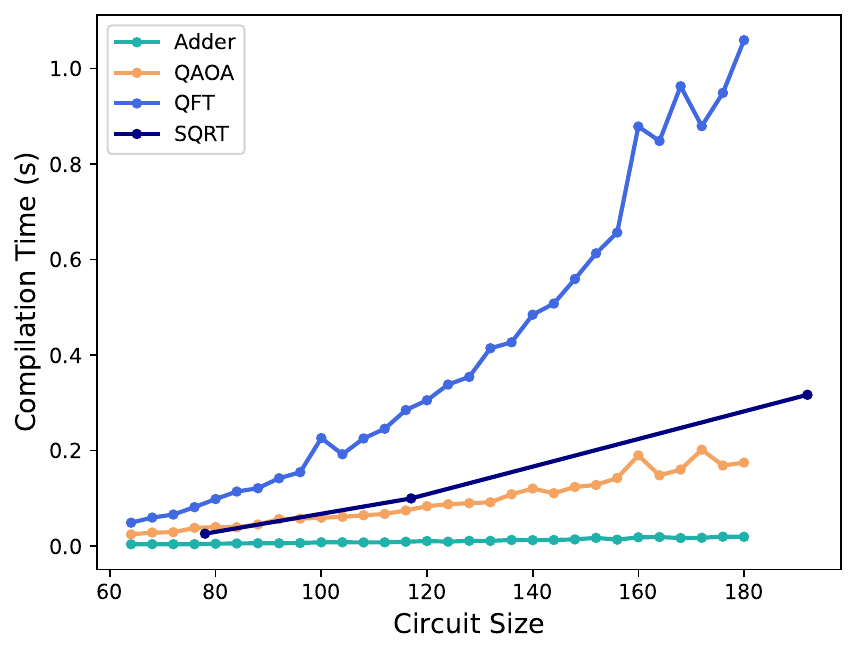}
    \caption{The compilation time varies with the size of the circuit. The circuit size refers to the number of qubits in the circuit, and the AOM size is 16.}
    
    \label{fig:scalability}
\end{figure}

Another distinct advantage of our method over existing approaches is its scalability. The computational complexity of our method is $O(ng)$, which ensures rapid circuit compilation even as the size of the circuit increases. To further demonstrate its scalability, we conducted a time analysis across various circuit sizes. The simulation results are depicted in Fig. \ref{fig:scalability}, where we selected four benchmarks and set the AOM size to 16.

The results indicate that the computational complexity largely aligns with our initial expectation of $O(ng)$.
Notably, the number of gates for both the adder and the QAOA circuits exhibits a linear relationship with the number of qubits. Furthermore, these gates are predominantly nearest-neighbor (or short-distance) gates. Consequently, the compilation time demonstrates a linear increase with the size of the circuit. 
As for the QFT circuit, it is observed that the number of gates increases quadratically with the number of qubits. 
As a result, the required compilation time increases approximately quadratically as the qubit count increases. These results demonstrate a trend that is consistent with expectations of computational complexity. 
Our method, however, demonstrates exceptional scalability. Even for the compilation of a QFT circuit with 180 qubits, the compilation time required by our approach does not exceed 1.2 seconds. In contrast, the previous method~\cite{wu2021tilt_iontrap} required more than 60 seconds to compile a QFT circuit with merely 64 qubits. This stark difference highlights the superior efficiency and scalability of our method in handling complex quantum circuits, even as the number of qubits grows.
%%%%%%%%%%%%%%%%%%%%%%%%%%%%%%%%%%%%%%%%%%%%%%%%%%%%%%%%%%%%%%%%%%%%%%%%%%%
\section{Related Work}
Numerous efforts have been devoted to the development of compilers on different quantum hardware. Compilers for superconducting quantum devices have been a major focus in recent research, primarily aimed at reducing swap gate insertions~\cite{li2019tackling,oddi2018greedy, huang2022reinforcement,zhu2023variation}. 
However, applying these methods directly to ion trap systems may not be ideal, as they are not tailored to the unique requirements of these devices.
These techniques model hardware constraints as a connected graph, where two-qubit gates can only be applied if an edge exists between any two vertices. Throughout this process, the graph's structure remains static and only the mapping of logical qubits to nodes is altered using SWAP gates.
In contrast, the TILT architecture introduces a changing topology during execution. Gate operations are restricted to an execution zone as defined by the TILT structure, resulting in a connected subgraph with a number of isolated vertices. In this configuration, it is not feasible to remap the isolated vertices through SWAP gates.
Each shuttle operation alters the composition of the connected subgraph, introducing a dynamic interaction between the shuttle process and the graph's structural integrity. This continuous modification increases the complexity of quantum algorithm implementation within the TILT framework, necessitating the recalculation of strategies to accommodate the graph's dynamic behavior.
As a result, traditional superconducting compiling algorithms are inadequate for handling the compilation process on TILT architectures.

While general compilers such as Qiskit \cite{aleksandrowicz2019qiskit} and t$\ket{\text{ket}}$ \cite{sivarajah2020t} compilers offer compilation capabilities for ion trap devices, it is important to note that their algorithms may not be suitable for varying ion trap system configurations and simulations. This mismatch could lead to sub-optimal results, as highlighted in the study by~\cite{wu2021tilt_iontrap}, which underscores the need for device-specific compilation strategies in ion trap quantum computing.

To best incorporate the properties of ion trap systems, there are only a few works focusing on quantum charge-coupled devices (QCCD) as well as linear devices. Previous studies, such as~\cite{Linke2017}, have focused on evaluating the performance of small-scale quantum devices. This particular study provides a comparative analysis between a 5-qubit ion trap device and superconducting quantum computing devices.  In our paper, we utilize real device parameters to estimate the properties and performance of ion trap systems as they scale up. This approach ensures a more accurate and tailored analysis for larger, more complex ion trap quantum computing devices.   

A mathematical formulation for obtaining the optimal solution using the integer linear programming approach was proposed in~\cite{wu2019ilp_iontrap}. However, as the number of qubits increases, the method does not scale well due to the substantial increase in time required. In~\cite{wu2021tilt_iontrap}, a heuristic algorithm was proposed to reduce the number of shuttles. In our study, we introduce a more efficient algorithm to optimize the shuttling number in ion trap devices. Our approach also takes into account the sympathetic cooling time and variations in gate implementations. This comprehensive consideration allows for a more accurate fidelity estimation, particularly in modern ion trap systems.

Besides linear-type ion trap compilers, other works have focused on QCCD architecture using heuristic functions~\cite{ murali2020architecting_iontrap, saki2022muzzle_iontrap, upadhyay2022_iontrap}.  They studied the use
of simulation techniques to study the impact of trap sizes,
topology, and gate implementations. Previous works, such as~\cite{Sargaran2019} and~\cite{Monroe2014}, have explored the potential of using reconfigurable and photonic interconnects for scalable trapped-ion systems, aiming to extend the architecture to thousands of qubits with fault-tolerant error correction capabilities. 
However, while these approaches hold promise for long-term scalability, they are unlikely to be feasible in the near future. In contrast, our work focuses on providing a more practical implementation strategy for near-term ion trap devices, addressing immediate scalability and efficiency challenges.

From an algorithmic perspective, our approach shares conceptual similarities with ideas in the pulse-level optimization domain~\cite{cheng2024epoc, smith2022programming, nguyen2021enabling, chen2013ultrasonic, gokhale2020optimized} and gate or circuit approximation~\cite{camps2020approximate, patel2021robust, patel2022quest}. 
A significant innovation in our manuscript is the introduction of a blocking algorithm specifically tailored for TILT compilation tasks. This algorithm not only reduces the number of shuttles required but also suggests potential avenues for further improvements.

%%%%%%%%%%%%%%%%%%%%%%%%%%%%%%%%%%%%%%%%%%%%%%%%%%%%%%%%%%%%
\section{Conclusion}\label{sec:conclusion}
In this paper, we address the optimization of shuttling numbers in ion trap systems, a critical factor for enhancing the efficacy of quantum computations. Our focus is on developing blocking algorithms that streamline the movement of ions, thus reducing operational time and mitigating error possibilities. By optimizing shuttling processes, we achieve better computational efficiency and reliability. 

Towards building fault-tolerant quantum computing (FTQC), it is typically requires many qubits with high fidelity and good connectivity. However, in the trapped-ion systems, simply increasing the number of ions in a single trap cannot simultaneously improve the quality and quantity of qubits. QCCD and transport of ions in linear tape architectures provide an excellent balance and provide a promising direction for realizing fault-tolerant quantum computing~\cite{QCCDQEC,QCCDQEC1}. The realization of this system depends on linking several linear-tape trapped-ion architectures, which necessitates the efficient compilation techniques of TILT. Our approach is poised to become a fundamental component in executing real quantum applications on large-scale QCCD devices.

Our current algorithm primarily relies on greedy methods, but there is potential to incorporate heuristic functions into this framework. This concept aligns with the qubit mapping of superconducting devices, albeit it may incur additional time overhead, it holds promise for finding improved solutions. In future work, it would be beneficial to integrate more sophisticated heuristic functions to further reduce the number of shuttling operations in our framework. Additionally, we have observed that insert swap operations result in decreased fidelity, therefore, in future work, efforts can be directed towards minimizing the use of insert swap gates. Drawing from the work on superconducting systems~\cite{li2019tackling, huang2022reinforcement,zhang2021time, yu2023symmetrybased}, we can also adapt their methods to our TILT cases. 

%%%%%%% -- PAPER CONTENT ENDS -- %%%%%%%%
\section*{ACKNOWLEDGMENTS}
We thank the anonymous reviewers for their helpful comments.
This work was partially supported by the National Key R\&D Program of China (Grant No.~2024YFE0102500), the Guangdong Provincial Quantum Science Strategic Initiative (Grant No.~GDZX2403008, GDZX2303007, GDZX2403001), the Guangdong Provincial Key Lab of Integrated Communication, Sensing and Computation for Ubiquitous Internet of Things (Grant No.~2023B1212010007), the Quantum Science Center of Guangdong-Hong Kong-Macao Greater Bay Area, and the Education Bureau of Guangzhou Municipality.
%%%%%%%%% -- BIB STYLE AND FILE -- %%%%%%%%
\bibliographystyle{IEEEtranS}
\bibliography{refs}
%%%%%%%%%%%%%%%%%%%%%%%%%%%%%%%%%%%%

\end{document}